\documentclass[floatfix,reprint,amssymb,aps, prd]{revtex4-1}
\usepackage{enumitem}
\usepackage[utf8]{inputenc}
\usepackage{amsmath}
\usepackage{lineno}
\usepackage{graphicx} 
\usepackage[caption=false]{subfig}
\usepackage{url}

\begin{document}

\title{Studies of an air-shower imaging system for the detection of
ultrahigh-energy neutrinos}

\author{Adam Nepomuk Otte}
\email{otte@gatech.edu} 
\affiliation{School of Physics \& Center for Relativistic
Astrophysics, Georgia Institute of Technology 837 State Street NW, Atlanta,
Georgia 30332-0430, USA  }
\email{otte@gatech.edu}
\date{\today}
\begin{abstract} 
We discuss the acceptance and sensitivity of a small air-shower imaging system
to detect earth-skimming ultrahigh-energy tau neutrinos. The instrument we study is
located on top of a mountain and has an azimuthal field of view of $360^\circ$. We find that the acceptance
and sensitivity of such a system is close to maximal if it is located about
2\,km above ground, has a vertical field of view of $5^\circ$, allows the
reconstruction of an at least $0.3^\circ$ long air-shower image, and features an
effective light-collection area of $10$\,m$^2$ in any direction. After three years
of operation, an imaging system with these features achieves an all-flavor
neutrino flux sensitivity of $5\times10^{-9}$\,GeV\,cm$^{-2}$\,s$^{-1}$\,sr$^{-1}$  at
$2\times10^8$\,GeV. 
\end{abstract}


\maketitle

\section{Introduction} 

The window to high-energy (HE) neutrino astronomy was recently opened by the
IceCube Collaboration with the detection of an astrophysical neutrino flux
\cite{Aartsen2013}, and evidence that supports the idea of blazars being a
source of neutrinos \cite{IceCubeCollaboration2018a,IceCubeCollaboration2018b}.
The latter conclusion comes from a correlation between the arrival time of
IceCube detected neutrinos and the observation of a flare in the electromagnetic
band, which highlights the importance of the multimessenger approach in modern
astrophysics. The detection of HE neutrinos raises a number of interesting
questions that are potentially answered by neutrino observations at higher
energies: What are the astrophysical sources of these neutrinos (see
\cite{Waxman1997,Meszaros2001,Loeb2006,Senno2016} and references in
\cite{Ahlers2018})? Are blazars really a source of neutrinos? What other sources
of HE neutrinos are there? These questions can be addressed by extending the
spectral measurements of the astrophysical neutrino flux to higher energies and
by reconstructing the arrival direction of neutrinos with better angular
resolution.

But, hunting down astrophysical neutrinos is not the only science case for
ultrahigh-energy (UHE, $>1$\,PeV) neutrino detectors; the long-standing quest
to understand the composition and the sources of UHE cosmic rays is another one. The connection
between neutrinos and  cosmic rays is made when UHE cosmic rays interact
with cosmic microwave background photons. The number of neutrinos produced in
these interactions depends on the composition of the cosmic rays
\cite{Kotera2010}.  While recent measurements with cosmic-ray experiments favor a heavier composition
\cite{Kampert2012,ThePierreAugerCollaboration2016,Abbasi2014,Yushkov2017,Aab2017a,Aab2017},
UHE neutrino observations would not only independently confirm these
findings but, more importantly, provide a unique handle to constrain source models
of UHE cosmic rays \cite{VanVliet2017}. Furthermore, would a combination of UHE
neutrino measurements with proton measurements obtained with next-generation UHE
cosmic-ray observatories (e.g. AugerPrime) allow one to simultaneously constrain the
UHE proton fraction and source evolution \cite{Moller2018}.

A third science case is tests of neutrino physics at the highest energies,
which could potentially hint at new physics beyond the standard model
\cite{GRANDCollaboration2018,Gorham2018a,Klein2013,Fox2018a,Ryabov2006}. One approach
would be to compare neutrino fluxes measured with experiments that are sensitive
to different neutrino flavors. In that context it is noteworthy that experiments
geared toward measuring extraterrestrial neutrinos have already provided
numerous important contributions to neutrino physics \cite{Cleveland1998,Fukuda1998,Ahmad2001,Ahmad2002,IceCubeCollaboration2017,Aartsen2018a}. 
Adding to the discovery potential for new physics is the detection of
neutrino candidates with ANITA, which have signatures expected from air showers but seem to contradict
the current understanding of neutrino physics \cite{Gorham2018a,Gorham2019}.

While the existence of $10^{9}\,$GeV neutrinos is noncontroversial and efforts
to hunt them are admirable, actually detecting them is a daunting task. The two
biggest challenges are the minute neutrino cross section and
the extremely low UHE neutrino flux. Both are overcome by instrumenting large detection volumes in a cost-effective
way.

At present, the efforts to detect UHE neutrinos can be divided into two
approaches. The first approach is to utilize, like IceCube or ANTARES/KM3Net
\cite{Ageron2011,Adrian-Martinez2016},
large volumes of ice or water, in which a neutrino interacts and a particle
shower develops. In fact, IceCube itself has sensitivity to UHE neutrinos and
the published IceCube limits are the presently most constraining ones up to
about $10^{11}$\,GeV \cite{Aartsen2018}. The ARA \cite{Allison2015,Allison2016}, ARIANNA
\cite{Barwick2015a,Barwick2017}, and ANITA Collaborations
\cite{Barwick2006,Gorham2016,Gorham2018} also use ice as a detection medium and
either deploy radio antennas
in ice (ARA, ARIANNA) or look for a particle shower induced radio
signature from a balloon (ANITA). For reviews about the detection of air showers
with radio and a more complete list of experiments that use or plan to use radio
for the detection of UHE neutrinos see \cite{Connolly2016,Schroder2017}. 

The second approach is the so-called earth-skimming technique. In the
earth-skimming technique, a UHE tau neutrino interacts within tens to hundreds
of kilometers inside Earth after entering it. The tau produced in the
interaction continues to propagate through Earth and emerges from the ground
if it does not decay before. If it emerges, the tau decays in the atmosphere
and initiates an air shower, which is then detected
\cite{Fargion1999,Fargion2001,Fargion2001a,Fargion2002,Feng2002,Fargion2006}. 

All major air-shower detecting astroparticle
experiments also search for tau neutrinos, even though these instruments are
designed to detect gamma rays or ultrahigh-energy cosmic rays and are not optimized
for tau-neutrino detection. For example, the Pierre Auger Collaboration has
published competitive limits above $10^8$\,GeV
\cite{Aab2015,ThePierreAugerCollaboration2017,PEDREIRA2018}.  

The most recent limit from a gamma-ray
Cherenkov telescope comes from the MAGIC Collaboration. They report a limit on
the diffuse neutrino flux, which was obtained by pointing the MAGIC telescopes
at the sea during times when gamma-ray observations could not take place due
to clouds \cite{Gaug2007,Ahnen2018a}.

Because these experiments are not optimized for earth-skimming neutrino
detections and in the case of pointed instruments like MAGIC have only a little
time to spare for neutrino searches, it is not surprising that several efforts
are ongoing to design and build dedicated earth-skimming neutrino detectors. 
One such project is GRAND \citep{Martineau2015,Bustamante2016,GRANDCollaboration2018}, which aims at the detection of radio
emission from tau initiated air showers by distributing several thousand
antennas over an about $100\,000$\,km${^2}$ large area in the Himalayas.

Another proposed experiment is NTA, which evolved from Ashra and uses the
air-shower imaging technique \cite{Sasaki2014,Hou2015}. For NTA, four optical
detector stations are proposed at Mauna Loa, Hawaii, of which three are located
on opposite mountains and the fourth one in between them.  One NTA detector station
consists of several units.  One unit is composed of four independent 1.5\,m
diameter mirrors each with one camera. The resolution of one camera is
$0.125^\circ$ \cite{Sasaki2014}.  Prototype detector stations have been built
and operated \cite{Asaoka2012}.  

The CHANT proposal discusses the deployment of a Cherenkov telescope system on a
long duration balloon flight or a satellite \cite{Neronov2016}. The  idea is now
pursued as part of SPB2 \cite{Adams2017} and POEMMA
\cite{Olinto2017}. Other experiments proposed in the past to image air showers
induced by taus are reported in \cite{Yeh2004,Liu2009,Cao2005}.

In this paper we discuss the acceptance and sensitivity of an air-shower imaging
detector, which is located on top of a mountain, which has an unobstructed
$360^\circ$ azimuthal field of view, and which uses the earth-skimming
technique. Our motivation is to understand (a) the sensitivity of such an
instrument to UHE neutrinos, (b) the impact of different detector parameters on
the sensitivity, and (c) the minimal detector configuration, which results in a
close to optimal sensitivity.

The main objective of the instrument would be to establish a diffuse UHE
neutrino flux. We thus do not go into a discussion of optimizing the
reconstruction of the energy and arrival direction of tau neutrinos. The
imaging technique, however, can provide energy resolution better than 20\% and
angular resolution better than $0.2^\circ$; see e.g.\,
\cite{Aleksic2016}.

We start the paper by introducing the experimental setup, i.e.\ the
topography assumed in the earth-skimming technique and the detector layout. We
then continue with the calculation of the joint probability that a tau neutrino
interacts inside Earth and a tau emerges from the ground. The calculation is
followed by a discussion of the consequences for the design of an air-shower
imaging experiment, which arise from the geometry of the air shower that
develops in the atmosphere. The implications of the air-shower geometry and the
emission pattern of the light from the air shower on the light intensity at the detector are discussed in the subsequent section followed by a
discussion of the acceptance and the sensitivity for different detector
configurations. Before closing with concluding
remarks, we briefly present design considerations for an actual instrument we
dub \emph{Trinity}.  

\section{Earth-skimming technique}

\label{sec:skimming}
\begin{figure}
  \includegraphics*[width=\columnwidth]{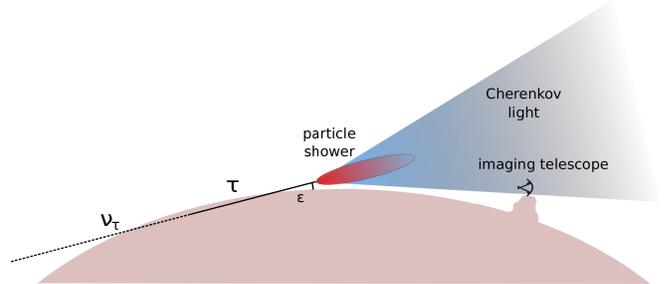}
\caption{Not-to-scale sketch of the earth-skimming technique and the emission
cone of Cherenkov light produced by charged shower particles. An imaging system,
receptive to Cherenkov and fluorescence light, is located on a
mountain and takes an image of the air shower by collecting some of the
light.  
  \label{fig:earthskimming}}
\end{figure}

Figure \ref{fig:earthskimming} shows the principle of the earth-skimming
technique. A tau neutrino enters Earth,
interacts and produces a tau. The tau propagates through Earth and, if it
has not yet decayed, emerges from the ground with an angle $\epsilon$,
which is the same angle under which the neutrino has entered Earth. For this
study we approximate Earth as a smooth sphere, i.e.\ without structure
and with a radius of 6371\,km.

If the tau decays in the atmosphere, a muon is produced in 17\% of the cases,
which traverses the atmosphere without further interaction and cannot be
detected.  Another 17\% of the decays produce electrons, which initiate
electromagnetic showers. The remaining 66\% of the decays result in charged and
neutral pions, which further decay into muons and gammas. The gammas initiate
electromagnetic showers, whereas the majority of muons leave the atmosphere
without further interaction.  The charged shower particles radiate about 0.1\%
of the shower energy as  Cherenkov radiation, which is beamed forward in a cone
around the shower axis. The photon intensity is approximately constant within
the cone out to an angle of about $1^{\circ}$ from the shower axis and
drops exponentially for larger angles.  We show later that the emission is
intense enough to be detected, even if the shower is several hundred
kilometers away. If the shower develops within tens of kilometers from the detector,
the shower can be imaged from all angles with the fluorescence light that is
emitted from nitrogen molecules in the atmosphere, which are excited by collisions with
shower particles.

The earth-skimming technique is only suitable for the detection of tau
neutrinos, even though the interaction cross section is about the same for all neutrino flavors in the
UHE band \cite{Cooper-Sarkar2011} and equal interaction rates are therefore expected
for all flavors. But electron neutrinos generate electrons, which
immediately produce a particle shower inside Earth. Muons produced in muon
neutrino interactions, on the other hand, emerge from the ground but the probability
for them to decay or interact within the field of view of a ground based
instrument and produce an air shower is negligible. While the muon itself emits
Cherenkov light, the intensity is so low that it would only be detectable if the
muon comes within a few hundred meters of the detector, which is unlikely. Only
taus decay and produce an air  shower within the field of view of a suitable instrument  with sufficiently high probability.  

The imaging of air showers is a very successful technique used
by a number of very high-energy gamma-ray instruments \cite{Archambault2017,Aharonian2006a,Aleksic2015a,Actis2011} and
ultrahigh-energy cosmic-ray experiments \cite{Abraham2010,Abbasi2015}. 
Sensitivity
determining factors of an imaging system are the following: 
\begin{enumerate}[label={\roman*)}]
\item The probability that a tau neutrino interacts on its path through
Earth and a tau emerges from the ground.
\item The probability that the tau decays and a particle shower 
develops within the field of view of the instrument.
\item The intensity of the Cherenkov/fluorescence light emitted in the direction
of the instrument is above the detection threshold of the instrument.
\item The ability to reconstruct the event with the recorded image and reject
the event if it is not a tau neutrino.
\end{enumerate}
The first three points determine the detector acceptance at trigger level
\begin{multline} \label{eq:1}
\Phi(E_\nu) = \\ \int_A \int_\Omega \int_{E_\tau} P_{\mbox{\footnotesize det}}(E_\tau)\cdot
P(E_\tau|E_\nu,\epsilon) \cdot
\sin(\epsilon)\,d E_\tau \,dA\,d\Omega
\end{multline}
where $P$ is the joint probability that a neutrino with energy $E_\nu$ interacts, a tau emerges from
the ground, and a particle shower is initiated. $P$ depends on the elevation
angle $\epsilon$ of the emerging tau and its energy
$E_\tau$. We discuss in the next section how we calculate $P$.
$P_{\mbox{\footnotesize det}}$ is the probability
that an air shower initiated by a tau with energy $E_\tau$ is detected. It
depends on the distance between the air shower and the detector, $\epsilon$, the
angle of the shower axis to the detector $\alpha$ (see later), and the detector
configuration. The acceptance is integrated over the plane $A$ defined by the
telescope location and the horizon in a given
direction. The elevation $\epsilon$ is measured relative to that plane.

We point out that the sensitivity of an earth-skimming
experiment is mostly determined by the topography of the area surrounding the detector. 
Throughout the paper, we approximate Earth as a smooth sphere, and the imaging
system is assumed to be located on a mountain, which provides an unobstructed
$360^\circ$ azimuthal field of view. We do not claim that this is the best possible
configuration. On the contrary, we view our choice as a baseline
configuration and are open to the idea that other topographies  result in better
sensitivities.

\section{Probability of Neutrino Interaction and Tau emergence}

We calculate the probability of a neutrino interacting and a tau emerging from
the ground with the parametrization from \cite{Dutta2005}, which includes a
treatment of the tau energy losses. But the parametrization does not include
neutrino regeneration  effects \cite{Alvarez-Muniz2018,Alvarez-Muniz2019}, and
our calculations thus underestimate the detectability of tau neutrinos below
$10^8$\,GeV.

We use Eq.\ (28) of \cite{Dutta2005} to calculate
the interaction probability of a tau
neutrino with energy $E_\nu$ and the emergence
probability of a tau with energy $E_{\tau}$ as
\begin{equation}
P = \frac{1}{\beta \rho E_\tau} \cdot P_\nu \cdot P_\tau
\end{equation}
where the neutrino interaction probability is
\begin{equation}
P_\nu = \sigma_{CC} \rho N_A \exp\left[  -D_\nu (\sigma_{CC} + \sigma_{NC})
\rho N_A  \right]
\end{equation}
and the probability for the tau to emerge from the ground is
\begin{equation}
P_\tau = \exp\left[- m_\tau/\left(\tau_\tau c \beta \rho E_\tau  \right)
 \cdot \left( 1 - e^{-\beta \rho D_{\tau}}
\right)\right],
\end{equation}
where $\rho=2.65\,$g/cm$^3$ is the density of rock and $\sigma_{CC}$ and $\sigma_{NC}$
are the energy dependent charged and neutral current neutrino cross sections from
\cite{Cooper-Sarkar2011}. $N_A$ is the Avogadro constant,
$\tau_\tau$ is the lifetime
of the tau, $m_{\tau}$ is its mass, and $c$ is the speed of light. $D_{\tau}$ is the
distance the tau has to travel through Earth before it emerges if its initial energy is
$0.8E_\nu$ and the energy when it emerges from the ground is supposed to be
$E_\tau$,
\begin{equation}
D_\tau = \ln\left[0.8 E_\nu/E_\tau\right]/(\beta\rho)\,.
\end{equation}
The distance $D_{\nu}$, which is the distance the neutrino travels through the
Earth before it interacts, is calculated from the combined trajectory $D$ of the tau
and the neutrino through Earth,
\begin{equation}
D = D_\nu + D_\tau\,\rightarrow\,D_\nu = D - D_\tau\,.
\end{equation}
For the energy loss $\beta$ of the tau while it propagates through
Earth we use Eq.\ (13) case III from \cite{Dutta2005}
\begin{equation}
\beta = \beta_0 + \beta_1 \ln(E/E_0)
\end{equation}
with $\beta_0 = 1.2 \times 10^{-6}\,\mbox{cm}^2/\mbox{g}$, $\beta_1 = 0.16 \times
10^{-6}\,\mbox{cm}^2/\mbox{g}$,
and $E_0 = 10^{10}$\,GeV.

\begin{figure}[!tb]
  \centering
  \includegraphics*[width=\columnwidth]{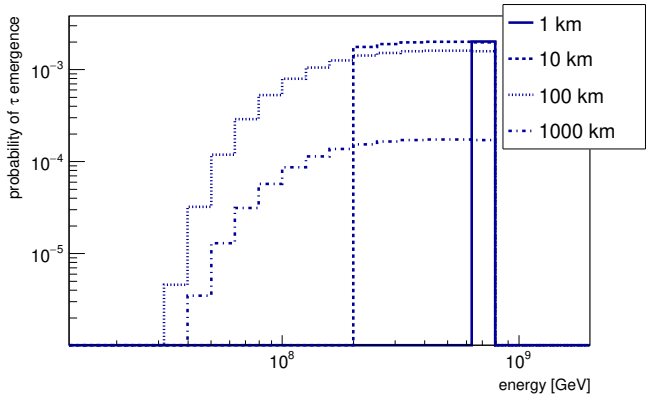}
\caption{Energy distribution of taus produced by $10^9$\,GeV tau neutrinos.
Distributions for targets with four different thicknesses are shown. The vertical axis gives the probability that the
neutrino interacts and a tau emerges from the target.
  \label{fig:energdistr}}
\end{figure}

As an example, we use the above parametrization to calculate the energy
distribution of taus that are produced by $10^9$\,GeV neutrinos and emerge from
the ground. We assume four targets with thicknesses: 1\,km, 10\,km,
100\,km, and 1,000\,km (see Fig.\ \ref{fig:energdistr}). The energy dispersion
is negligible for a 1\,km thick target and steadily increases for thicker
targets.  For a  100\,km target, the dispersion covers one decade of
energy. 

\begin{figure}[!tb]
  \centering
  \includegraphics*[width=\columnwidth]{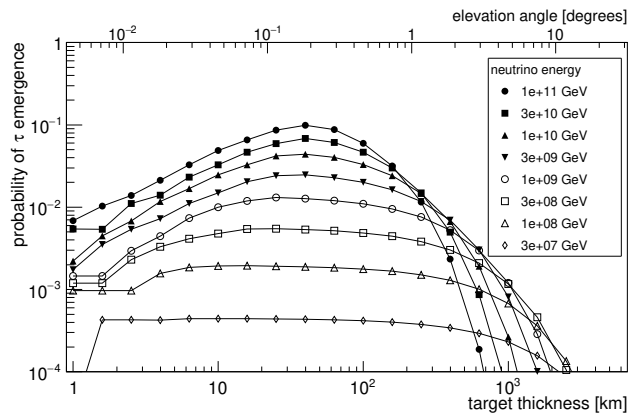}
\caption{Probability of a tau neutrino interacting and a tau emerging from a
target with density $\rho=$2.65\,g/cm$^3$ (rock). The top axis 
shows the elevation angle under which a neutrino has to enter Earth in order to traverse
the distance given on the bottom axis.
}
  \label{fig:emergprob}
\end{figure}

Integrating each of the distributions gives the probability that a $10^9$\,GeV
neutrino interacts and a tau emerges. This so-called emergence probability is
shown in Fig.\,\ref{fig:emergprob} for different neutrino energies as a
function of target thickness. For $10^9$\,GeV neutrinos the probability steadily
increases with target thickness and rapidly drops beyond 1,000\,km.  The axis on
top of the figure shows the elevation angle $\epsilon$ for which a neutrino
entering Earth would traverse the same distance as shown on the bottom axis
before reemerging from the ground \footnote{See figure \ref{fig:earthskimming}
for a definition of $\epsilon$.}.

For $>10^9$\,GeV neutrinos, the emergence probability is maximal if the target is 
between 10\,km and 500\,km thick, which corresponds to elevation angles between
$0.5^\circ$ and $2^\circ$. For thicker targets, the taus decay inside the
target and for thinner targets, the interaction probability goes down.
 For lower neutrino energies an optimal target thickness is less evident
because of the smaller neutrino interaction cross section and the shorter tau
lifetime.

\section{Geometrical Constraints in Air-Shower Imaging}

When a tau emerges from the ground, it decays and can induce a particle shower.
The particles of the shower produce Cherenkov and fluorescence light, which is
used to image the shower with an air-shower imaging instrument.

The Cherenkov and fluorescence light have very different characteristics.
Cherenkov emission is beamed into the direction of shower development, whereas
fluorescence emission is isotropic. Integrated over all emission angles, the
Cherenkov emission is orders of magnitude more intense than the fluorescence
emission. The Cherenkov spectrum is proportional to $1/\lambda^2$ at the point
of production, whereas the fluorescence light is emitted in a few narrow lines
around 340\,nm \cite{Ave2007,Ave2013}. The shape of the Cherenkov spectrum is
different at the detector due to scattering and absorption in the atmosphere. We
show simulated spectra later. Another striking difference is the arrival-time
distribution of the photons. The Cherenkov photons arrive within a few
nanoseconds if the shower is viewed head on and within several microseconds if
the shower is viewed under an angle of $30^\circ$. We give an example of a
simulated time distribution later. The fluorescence photons arrive over several
microseconds.  As a result of these different characteristics, showers that
develop more than $\sim50$\,km from the telescope can only be imaged with
Cherenkov light. The fluorescence light becomes the dominant
detection channel at closer distances. 

Even though both emission mechanisms have vastly different characteristics, the
requirements to record a shower image derive mostly from geometry and equally
apply to both detection channels:

\begin{itemize}
\item The shower image has to be fully contained in the field of view of the
telescope. 
\item Light from both ends of the air shower need to arrive at the
telescope with sufficient intensity. 
\item The image has to be of a minimal length in order to be reconstructable. 
\end{itemize}

\begin{figure*}[!htb]
  \includegraphics*[width=\columnwidth]{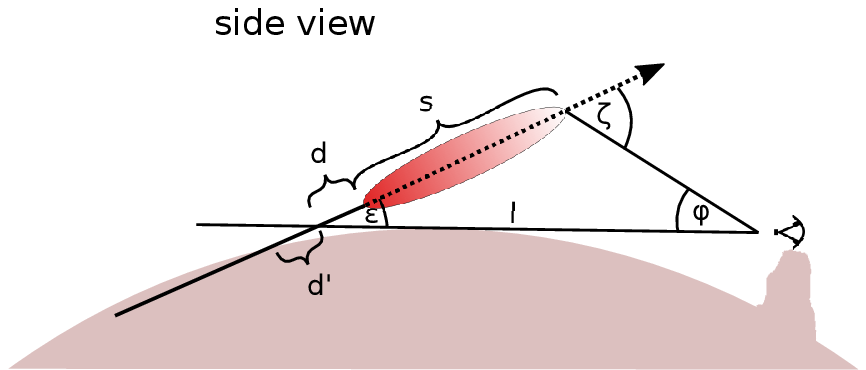}
\hspace{20mm}
  \includegraphics*[width=0.4\columnwidth]{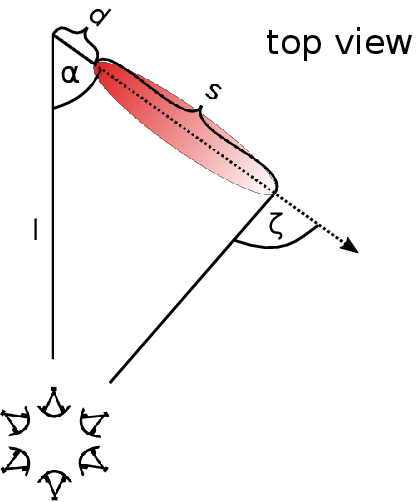}
\caption{ 
Side and top views of the air shower and the imaging system to illustrate
geometrical requirements, which have to be fulfilled to obtain a complete image
of the shower. Two of the conditions are (a) Cherenkov light from the tip of the shower
has to reach the telescope in sufficient intensity, which is constrained by the
maximum Cherenkov angle $\zeta_{\mbox{\footnotesize max}}$. (b) The shower needs to be fully
contained in the vertical field of view $\phi$ of the telescope. The remaining
conditions are discussed in the text. The angles and distances marked in the
figures are projections of their actual quantities with the exception of
$\alpha$, which is the angle between $l$ and the projection of the shower axis
onto the ground. $l$ is the distance between the telescope and the point where
the tau enters the field of view.  } \label{fig:geomview} \end{figure*}

Figure \ref{fig:geomview} shows how these requirements translate into
geometrical constraints. The particle shower is depicted in the figure in
relation to the imaging system with all relevant distances and angles. The
trajectory of the tau and the main shower axis are represented by the arrow. 

If the tau emerges below the horizon from the ground, as it is shown in Fig.\
\ref{fig:geomview}, it first propagates a distance $d'$ before it enters the field of view of the
telescope. The distance to the telescope when the tau enters the field of view is $l$. The fraction of taus that has
not decayed while propagating through $d'$ is $\exp\left(-d'/\lambda \right)$, where the decay
length is $\lambda = E_\tau \cdot c \cdot \tau_\tau / m_\tau = 49\,$km$\cdot
E_\tau/10^9$\,GeV.  
The decay length $\lambda$ solely depends on the tau energy $E_\tau$,
which is between 10\% and 100\% of the neutrino energy (see Fig.\
\ref{fig:energdistr}). For the neutrino energies of interest, $10^8$\,GeV to
$10^{10}$\,GeV, the tau energies are between $10^7\,$GeV and
$10^{10}\,$GeV. The distance $d_{90} =-\ln(0.1)\cdot\lambda$ over which 90\% of
the taus decay is then between 1\,km ($10^7\,$GeV) and 1,000\,km ($10^{10}\,$GeV).

After propagating an additional distance $d$ in the telescope field of view, the
tau decays and a particle shower is initiated, which develops over the distance
$s$. The necessity of imaging the entire shower requires that the angle $\phi$
has to be smaller than the field of view of the camera above the horizon and in
turn sets the maximum possible value for $d$. If more than 90\% of the taus
decay over $d$, we  readjust $d$ to $d_{90}$. This is to simplify the acceptance
calculation for which we would otherwise have to integrate along $d$ and
evaluate at each step in the integral if enough light reaches the detector.
Instead, we assume that all taus decay when reaching the end of $d$ and evaluate
only for that situation if enough light reaches the detector from the tip of the
shower, i.e.\ if the intensity emitted under angle $\zeta$ is above the
detection threshold of the telescope.  If the intensity is below the detection
threshold, we reduce $d$ and thus $\zeta$ until the
intensity is above threshold. We do this because the Cherenkov intensity
increases exponentially for smaller angles of $\zeta$ (see later).  If no value
for $d>0$ can be found that satisfies all requirements, the shower is considered
unobservable.

The length of the shower $s$ is estimated with the Heitler model for
electromagnetic showers \cite{Longair2011,Heitler1984}, which states that the
number of particles doubles and the energy per particle halves for every
radiation length $X_0$ due to pair creation of gammas and bremsstrahlung by
electrons/positrons.  The number of particles reaches its maximum when the
average energy per particle reaches the critical energy $E_c=88\,$MeV for
electrons in air \cite{Tanabashi2018}. In this model, the shower develops over
\begin{equation} X = \frac{\ln(E/E_c)}{\ln(2)} \end{equation} radiation lengths. 

For the estimate of the shower length, we assume that 50\% of the tau energy is
deposited in the electromagnetic part of the shower. For the tau energies of
interest (see above), the particle shower then develops over 27--37 radiation
lengths. If the shower develops close to the ground, which implies that the air
density is constant and one radiation length is 304\,m, the shower length $s$ is
8\,km--11\,km. 

A constant air density is a good approximation for the development of showers
initiated by $10^7$\,GeV taus. Even for an extreme tau emergence angle of
$10^\circ$ (see Fig.\ \ref{fig:emergprob}), the shower is fully developed at an
altitude of 1.6\,km. At that altitude the air density is still 80\% of the
density at sea level.  For the highest-energy taus, the emergence probability
(see Fig.\ \ref{fig:emergprob}) reaches 10\% of the peak probability when the
emergence angle is about $3^\circ$. For that angle the shower is fully developed
at an altitude of 600\,m above ground. 

But because the decay length is much longer for high-energy taus, the shower
develops higher up in the atmosphere and, therefore, over a longer distance
because of the lower air density. We nevertheless use the shorter shower length
calculated for an air density close to ground, which is conservative because it
results in an increased number of events that are rejected because they fail the
minimal-length requirement of the image in the reconstruction.

Summarizing, the detection probability of a tau, $P_{\mbox{\footnotesize det}}$
in Eq.\ (\ref{eq:1}), is evaluated by calculating the probability that the tau decays
inside the field of view of the camera under the constraint that the shower has
to be fully contained inside the camera field of view. A shower image is
considered reconstructable if its image length is above the required minimum and
the detected light intensity from both ends of the shower is above the detection
threshold of the telescope.  The decay length $d$ is adjusted to meet the
minimum intensity requirement or, if it is shorter, the distance over which 90\%
of the taus decay.

\section{Cherenkov Light Density}

Crucial for a valid calculation of the detection probability of a tau
$P_{\mbox{\footnotesize det}}$
is a good understanding of the amount of Cherenkov light and fluorescence light
reaching the detector. In this section, we discuss the Cherenkov light density
followed by the fluorescence light density in the next section.  

The amount of Cherenkov light, that reaches the telescope depends on the tau
elevation $\epsilon$, azimuth $\alpha$, distance to the shower $l$, and location
of the telescope above ground $h$ as defined in Fig.\ \ref{fig:geomview}. We
derive parametrizations of the Cherenkov light intensity at the detector for
four different telescope locations above ground (0\,km, 1\,km, 2\,km, and
3\,km). For the parametrization we have simulated air showers initiated by
$3\times10^4$\,GeV gamma rays. We use gamma rays because they initiate pure
electromagnetic showers and deposit all the energy into the shower, which for
our purpose yields a robust Cherenkov light intensity per tau energy deposited
in the shower (we assume that 50\% of the tau energy is deposited in an
electromagnetic shower).  It is, furthermore, sufficient to
simulate one energy because the Cherenkov intensity is proportional to the
shower energy and can thus easily be scaled in the calculation of the detection
probability to match the energy deposited in the shower. As a cross-check,
we also derived a parametrization by simulating $10^6$\,GeV gamma rays
and found both parametrizations to be in agreement.

The air showers are simulated with CORSIKA version 75000 \cite{Heck1998}, which
we modified to allow the simulation of Cherenkov photons up to 900\,nm.
CORSIKA does not simulate Cherenkov emission for upward going showers, which is
why we have configured CORSIKA with an atmosphere of constant density and
have simulated straight downward going showers. The gamma
rays are released at heights of 25\,km, 55\,km, 85\,km, 135\,km, and 195\,km.
The arrival time, wavelength, impact angle, and position of all Cherenkov
photons, that reach the ground are saved.

Atmospheric absorption of the Cherenkov photons is not simulated in CORSIKA but
subsequently applied. For the altitude and wavelength dependent extinction
coefficients in the atmosphere, we take the model that is used by the VERITAS
Collaboration in the analysis of winter season data. The model was simulated
with MODTRAN \cite{modtran}.  Mie scattering is considered to be an absorption
and not a scattering process in the model, which results in an underestimation
of the Cherenkov intensity outside of the primary Cherenkov cone ($>1^{\circ}$)
by a few percent \cite{Neronov2016,Louedec2014,Giller2012}.  Refraction in the
lower atmosphere, which results in a bending of the photon trajectory along the
Earth's surface and an increase in photon intensity at the telescope, is also
not taken into account.

The absorption is calculated by translating and rotating the CORSIKA simulated shower 
such that (a) the shower axis points upward with an
elevation angle $\epsilon$ (see Fig.\ \ref{fig:geomview} \footnote{Note that
unlike what is depicted, we assume a flat Earth surface for the absorption
calculation}), (b) $d=5$\,km, and (c) the plane in which all the photons are recorded
in CORSIKA intersects with the location of the telescope. Each photon in the
plane is then rotated around the shower axis and the telescope is moved parallel
to the ground such that the positions of the photon and the telescope coincide.
The photon is then ray traced back to its origin and the
absorption applied in steps of 1\,km.

\begin{figure}[!tb]
  \centering

  \includegraphics*[width=\columnwidth]{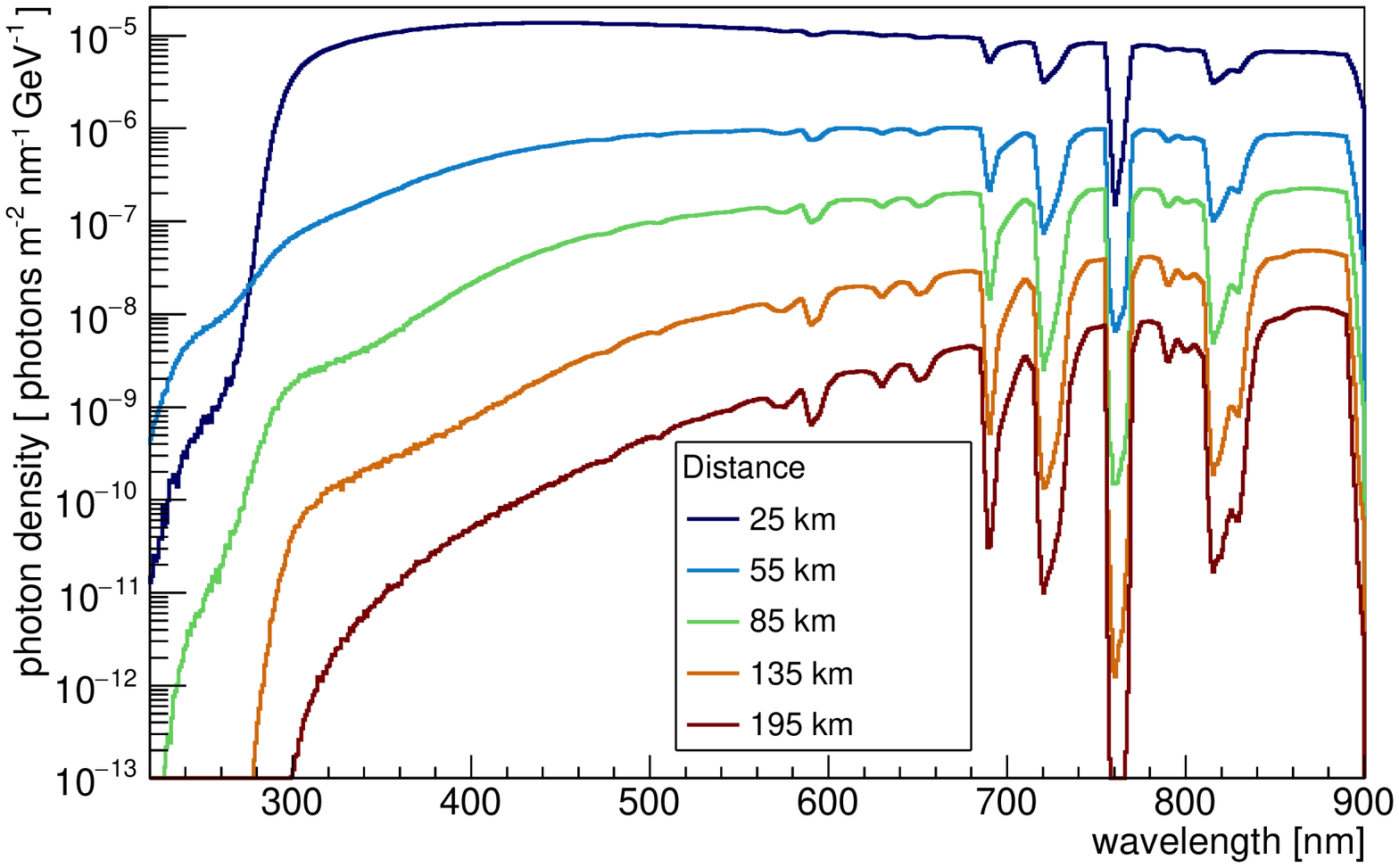}
\caption{Cherenkov photon spectra at the detector for showers starting
 in distances of 25\,km, 55\,km, 85\,km, 135\,km, and 195\,km from the detector.} 
\label{fig:CSpectAbsorbed} \end{figure}

After absorption is applied, simulated Cherenkov spectra at the location of the
telescope such as the ones in Fig.\ \ref{fig:CSpectAbsorbed} are obtained. In this
particular case the telescope is located 1\,km above the ground, the elevation
$\epsilon$ is $1^{\circ}$, and $\alpha$ is $0.25^{\circ}$. It is evident how the
peak of the detected spectrum shifts toward longer a wavelengths with an increasing distance
between the telescope and the shower, from 400\,nm at 25\,km to 900\,nm at
195\,km.

\begin{figure}[!tb]
  \centering
  \includegraphics*[width=\columnwidth]{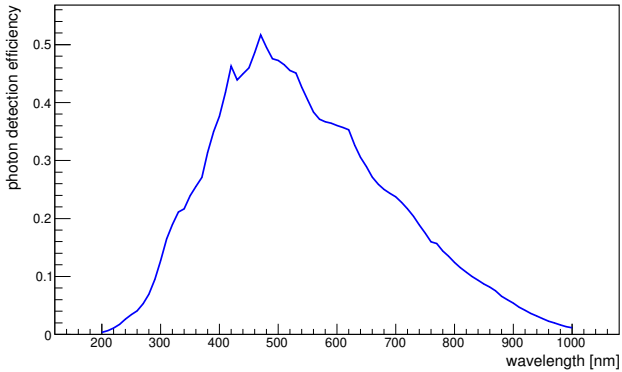}
\caption{Photon detection efficiency of the S14520-6050CN SiPM from Hamamatsu
for a bias, which yields a 90\% breakdown probability at 400\,nm.
}
  \label{fig:PDE}
\end{figure}

The spectral response of silicon photomultipliers (SiPMs) is a good match to the
heavily absorbed Cherenkov spectra. Reviews of SiPMs can be found in
\cite{Renker2009,Otte2015a}. For this study, we adopt the photon detection
efficiency (PDE) of the Hamamatsu SiPM S14520-6050CN when it is operated at a
bias voltage that yields a 90\% breakdown probability at 400\,nm (see Fig.\
\ref{fig:PDE}).  The PDE measurement was done with the setup described in
\cite{Otte2017}.  The PDE peaks at 470\,nm with 51\%, the full width at half
maximum of the spectral response spans from 360\,nm to 680\,nm, and a long tail
of the PDE extends to 1000\,nm. With this spectral response about 30\% of the
Cherenkov photons that arrive at the telescope are detected. Other
characteristics of the S14520-6050CN are an optical cross talk of less than 1.5\%
and 50$\,\mu$m size cells \cite{Otte2018}.

\begin{figure}[!tb]
  \centering
  \includegraphics*[width=\columnwidth]{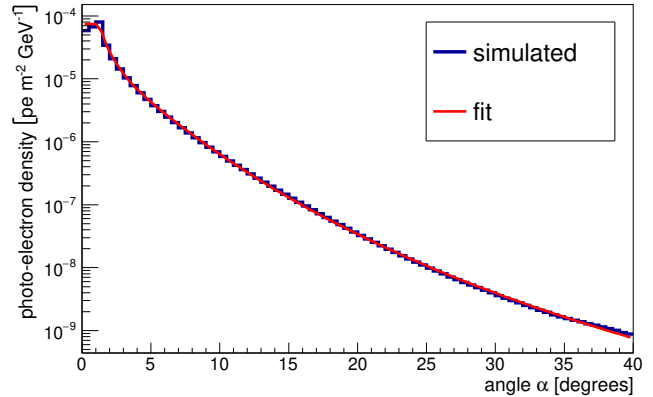}
\caption{
Simulated (blue line) and parameterized (red line) photoelectron densities per GeV shower
energy as a function
of angle $\alpha$. The distance to the shower is 55\,km, the shower elevation
angle is $0^\circ$, and the telescope is located on the ground ($h=0$\,km).
}
  \label{fig:masterradialdistr}
\end{figure}

Integrating the product of the simulated Cherenkov photon density and the PDE
and dividing the integral by the simulated gamma-ray energy, we obtain the
density of the detected Cherenkov photons (photoelectrons or pe) per GeV shower
energy.  Figure \ref{fig:masterradialdistr} shows the photoelectron density per
GeV air-shower energy as a function of $\alpha$ for an air shower with an
elevation of $0^\circ$ and a starting point 55\,km away from the telescope.
The telescope is located on the ground $h=0$\,km.  The density is
roughly constant out to $1^\circ$, which is not evident from the coarse binning
of $0.5^\circ$ but expected \cite{Nerling2006}.  Above  $1.3^\circ$, the
intensity is well described with two exponential functions 

\begin{multline}
\rho(\alpha') =
3.32\cdot10^{-4}\,e^{-\frac{\alpha'}{0.581^{\circ}}}\frac{\mbox{\small
pe}}{\mbox{\small m}^2\,\mbox{\small GeV}}\\
+4.26\cdot10^{-5}e^{-\frac{\alpha'}{1.95^{\circ}+0.0427\alpha'}}\frac{\mbox{\small
pe}}{\mbox{\small m}^2\,\mbox{\small GeV}}.
\end{multline}

Note that $\alpha'$ is the angle between the shower axis and $l$. In this
particular case ($\epsilon=0^\circ$ and $h=0$\,km), $\alpha'$  coincides with $\alpha$,
which is defined as the projection of $\alpha'$ onto the plane defined in Fig.\
\ref{fig:geomview}. 

The remaining density distributions for distances of
25\,km, 55\,km, 85\,km,\ 135\,km, and 195\,km and elevations from $0^\circ$ to
$10^{\circ}$ are well described by functions of the form
\begin{equation}\label{eq:densityfull}
N_0\cdot\rho(\alpha')\left(2-e^{-\frac{\epsilon}{\lambda_0}}\right)\cdot
e^{-\frac{l-55\,\mbox{\tiny km}}{\lambda_1+(l-55\,\mbox{\tiny
km})\cdot\lambda_2}}\quad,
\end{equation}
where we use $\rho(\alpha')$ from the previous equation. Angle $\alpha'$ is
again the angle between the shower axis and $l$ and is
calculated as $\alpha'=\sqrt{\alpha^2+(\epsilon-\tan^{-1}(h/l))^2}$.  The best
fit parameters of the photoelectron density function are listed in Table \ref{tab:denspars}
for four different telescope heights $h$.

\begin{figure}[!tb]
  \centering
  \includegraphics*[width=\columnwidth]{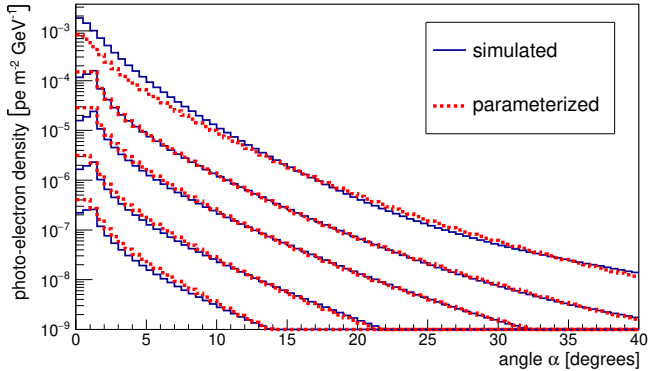}
\caption{
Simulated (blue line) and parameterized (red line) photoelectron density distributions
for a  $1^\circ$ shower elevation and a telescope location above ground of 1\,km. The
topmost distribution is for a 25\,km distance to the shower. The distributions
shown below it correspond --- from top to bottom --- to distances of 55\,km, 85\,km, 135\,km, and 195\,km.
}
  \label{fig:PEDensity}
\end{figure}

Figure \ref{fig:PEDensity} shows, as an example, simulated and parameterized
photoelectron density distributions for a telescope located 1\,km above ground
and a shower elevation of $1^{\circ}$. Discrepancies between the
parameterized and simulated densities are evident at small angles $\alpha$, while
the agreement is good for large angles. Even though the discrepancy is as large
as a factor of 2 at some angles, the impact on the acceptance calculation is
tolerable.

The discrepancies are tolerable because the density distributions are only used
to determine the integration limits for $\alpha$ in the acceptance calculation,
i.e.\ at what value of $\alpha$ the photoelectron density drops below
the detection threshold of the telescope.  The limits for $\alpha$
derived from the parametrization deviate by no more than  $0.5^\circ$ from the
simulated value if the distance to the shower is more than 25\,km. Such a small
discrepancy does not have a significant impact on the acceptance calculations.
At a distance of 25\,km, the parametrization deviates from the simulated value
by as much as $1.5^\circ$. This is again acceptable because the threshold for
the minimal photoelectron density is reached at much larger $\alpha$, where the
difference between simulation and parametrization is again acceptable.
Furthermore, Cherenkov detected events do not contribute much to the acceptance
if they originate within a few tens of kilometers from the telescope, which we show
later.

Besides showing that the parametrization is a good fit to the simulated
distribution, Fig.\ \ref{fig:PEDensity} also demonstrates that air showers can
be detected out to large angles of $\alpha$. For example, according to the
figure, a shower with $10^8$\,GeV can be detected from a distance of 195\,km out
to an angle $\alpha=5^\circ$ with a 10\,m$^2$ telescope and a 24\,photoelectron
threshold. If the shower is 135\,km away, the angle increases to $20^\circ$.

\begin{table}[htb]
\caption{\label{tab:denspars} Best fit parameters for Eq.\
(\ref{eq:densityfull}) for four different telescope heights above
the ground.
}
\begin{tabular}{ccccc}
\hline\hline
\begin{tabular}[c]{@{}c@{}}Height \\ (km)\end{tabular} &
\multicolumn{1}{c}{\begin{tabular}[c]{@{}c@{}}$N_0$\\ (.)\end{tabular}} &
\multicolumn{1}{c}{\begin{tabular}[c]{@{}c@{}}$\lambda_0$\\ (degrees
$10^{-4}$)\end{tabular}} & \begin{tabular}[c]{@{}c@{}}$\lambda_1$\\
(km)\end{tabular} & \begin{tabular}[c]{@{}c@{}}$\lambda_2$\\
($10^{-2}$)\end{tabular}\\ \hline
0 & 1 & 1.64 & 16.7 & 5.09 \\
1 & 1.03 & 1.91 & 16.6 & 5.23 \\
2 & 1.53 & 7.12 & 18.0 & 5.95 \\
3 & 2.37 & 51.3 & 19.0 & 6.42 \\\hline\hline
\end{tabular}
\end{table}

\subsection{Arrival-time distribution of Cherenkov photons}

\begin{figure}[!tb]
  \centering
  \includegraphics*[width=\columnwidth]{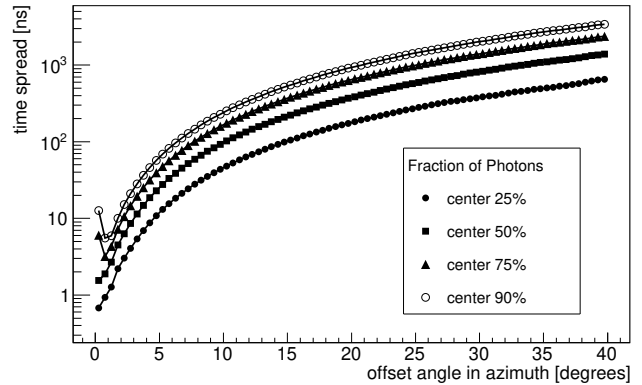}
\caption{Time during which Cherenkov photons arrive at the detector as a function of angle $\alpha$.
Each curve represents a time period during which a different fraction of the
Cherenkov photons arrives at the telescope. All intervals are centered at the median photon arrival time.
}
  \label{fig:ArrivalTimeSpreadvsAzimuth}
\end{figure}

The time during which the Cherenkov photons arrive at the detector depends on
the angle $\alpha$ and is shown in Fig.\ \ref{fig:ArrivalTimeSpreadvsAzimuth}
for 30\,TeV electromagnetic showers starting at 130\,km from the telescope with an
elevation angle of $0.3^{\circ}$.  That elevation and distance is typical for
tau-neutrino events, which contribute to the peak in the radial acceptance (see
later sections). The telescope is located 2\,km above the ground.  Each point in
the figure shows the time period centered at the median photon arrival time
during which a certain fraction of Cherenkov photons arrives at the telescope if
the shower is viewed under an angle $\alpha$. The fractions are 25\%, 50\%, 75\%,
and 90\%.

If $\alpha$ is less than $3^{\circ}$, 50\% of the Cherenkov photons arrive at
the detector within 10\,ns. If $\alpha$ is $10^{\circ}$, 50\% of the photons
arrive within 100\,ns and if $\alpha$ is $40^{\circ}$, the same fraction of
photons arrives within $\sim1\,\mu$s.

\section{Fluorescence Light Density}

Calculating the intensity of the fluorescence emission at the telescope is much
simpler than for the Cherenkov emission because the emission is isotropic and
concentrated to within a 130\,nm wide wavelength band. 

The absolute intensity of the 337\,nm fluorescence line was measured by the
AIRFLY Collaboration \cite{Ave2013}. We use their measurements in our
calculations but neglect variations of the intensity due to changing air
pressure and temperature, and we use an average value of 6,000 photons per GeV
shower energy. The same team also measured the relative intensities of the
fluorescence lines between  292\,nm and 428\,nm \cite{Ave2007}. We multiplied
the relative intensity of each line with the PDE of the S14520-6050CN SiPM and
then scaled them to the absolute intensity of the 337\,nm line and arrive at an
integral photoelectron intensity of 5,960 photoelectrons per GeV shower
energy. The emission is, furthermore, attenuated in our calculations with an
attenuation length of 9.5\,km, which is the attenuation length from the VERITAS
atmospheric absorption model at these wavelengths.

\section{Trigger Threshold}
The detectability of an air shower depends not only on the amount of light that
reaches the telescope but also on the minimum amount of light needed to trigger
the readout electronics of the telescope, which we derive in this section.  In
imaging systems, a trigger decision is commonly derived by constantly monitoring
the signal amplitudes of all pixels in the camera. If the signal amplitude of
one pixel goes above a predetermined threshold that pixel is said to be
triggered. But the command to read out the camera is only sent if a certain
topology appears among the triggered pixels. In our study that topology consists
of two pixels that trigger within a 10\,ns window and are located
next to each other in the camera. 

In the absence of air showers, the signal of a pixel is due to electronic noise and
detected background photons. SiPMs are single-photon detectors with high
intrinsic gain, and the signals of individual photons, i.e.\
photoelectrons, are clearly identifiable in a properly designed signal chain.
The trigger threshold of a pixel is thus not determined by the electronic noise but
by the intensity of the photon background. This so-called night-sky background
(NSB) is due to zodiacal light, airglow, star light, and artificial sources
(see e.g.\ \cite{Leinert1998}). Due to the randomness of the photon arrival
times, it can happen that the signals of photons pile up to
a signal that triggers the telescope readout. The rate by which that happens
depends on the intensity of the NSB, the pixel trigger threshold, the shape of the
photoelectron signals, and the coincidence window. For this study, the
pixel trigger threshold is set such that the expected telescope trigger rate due to
NSB fluctuations is 1\,Hz.

We measured the NSB with the S14520-6050CN at the Whipple Observatory in Arizona
at an altitude of about 1\,km above the surrounding area. Pointing the SiPM at
the zenith during a clear and moonless night, we detected an NSB rate of
$3.7\times10^6$\,photoelectrons/s/mm$^2$/sr when the SiPM is biased at the same voltage
we used for the PDE measurement shown in Fig.\ \ref{fig:PDE}. It is expected
that the NSB rate increases by a factor of 4 when the sensor is pointed at a
zenith angle of $80^\circ$ and that it drops again sharply for larger zenith angles
because of increased scattering and attenuation in the atmosphere
\cite{Leinert1998}. It is furthermore expected that, when pointing below the
horizon, the detected NSB rate should be close to the rate detected when pointing at
the zenith. Instead, we measured a 40 times higher NSB rate when pointing at the
horizon, which was into the direction of Tucson, a city with a population of
500,000 that is 50\,km away. The measurement at the horizon is thus not
representative for a dark site, and we instead use the measurement at the zenith to
derive pixel trigger thresholds for our acceptance and sensitivity calculations. 

In order to determine the pixel trigger threshold, which results in a telescope
trigger rate of 1\,Hz, we first calculate the expected NSB rate $R_D$  in a
camera pixel. For this we assume a telescope with the wide field-of-view optics
developed for MACHETE \cite{Cortina2016}. A MACHETE optics with a 1\,m$^2$
effective mirror area and a camera with an angular resolution of $0.3^\circ$
results in pixels with an area of 36\,mm$^2$ and an angular acceptance of 0.842\,sr.
Multiplying the measured NSB rate with these values yields an
$R_D=10^8$\,counts/s per square meter effective mirror area. The dark count rate
scales proportional to the area of the telescope mirror.

The rate of $n$ photoelectrons piling up depends on $R_D$ and the photoelectron signal
shape. Here we adopt an effective signal width of $\Delta t = 10\,$ns, which is
readily achievable, and assume that all photons detected within
10\,ns line up perfectly.  The rate $R_n$ of $n$
photoelectron signals piling up is then calculated as 
\begin{equation}
 R_n = R_D  \left[ 1 - \sum_{i=0}^{n-1} \frac{\mu^ie^{-\mu}}{i!} \right]
\end{equation}
where $\mu=R_D\Delta t$ is the average number of NSB signals in $\Delta t$. 

Finally, we calculate the telescope trigger rate for a given $R_n$ by assuming a
camera with a field of view of $5^\circ\times60^\circ$, i.e.\ 3,333
pixels, and the
aforementioned two next neighbor coincidence logic with a 10\,ns coincidence
window. $R_n$ is then increased until the telescope trigger rate drops down to 1\,Hz.  For
telescopes with effective mirror sizes of 1\,m$^2$, 5\,m$^2$, 10\,m$^2$, and
100\,m$^2$ the thus derived pixel trigger thresholds are 10\,pe, 22\,pe, 24\,pe,
and 155\,pe, respectively.  

In the acceptance calculation we mimic a twofold coincidence trigger by
requiring that the total number of detected Cherenkov photons is at least twice
the single pixel threshold value. The acceptance does not depend strongly on the
trigger threshold because the Cherenkov photoelectron density drops
exponentially with $\alpha$ and the distance from the telescope, which we have shown
in previous sections. The biggest caveat
of our simplified trigger is that it does not account for the arrival-time
distribution of the Cherenkov photons, which is a few nanoseconds for small
$\alpha$ angles but can reach several hundred nanoseconds for large $\alpha$
angles (see Fig.\ \ref{fig:ArrivalTimeSpreadvsAzimuth}).

\section{Acceptance Studies}

In this section we combine our previous findings and calculate the integral
acceptance for $10^{8.5}$\,GeV to $10^{9.5}$\,GeV tau neutrinos as a function of
distance from the telescope.  We assume that the neutrino flux follows a power
law with index $-2$ and correspondingly weight the acceptance when integrating
over neutrino energy.  We do separate acceptance
calculations for events detected via fluorescence emission and for those
detected via Cherenkov emission. 

Our baseline telescope is located 2\,km above ground, has a $5^\circ$ vertical
field of view of which $2^\circ$ point above the horizon and $3^\circ$ point
below the horizon, and has a $10\,$m$^2$ effective mirror area anywhere in the
telescope field of view. The minimum required image length in the reconstruction
is $0.3^\circ$.  We study how changing each of these characteristics impacts the
acceptance and, in doing so, show that our baseline is close to the telescope
configuration, which yields the best possible sensitivity.

\begin{figure}[!tb]
  \centering
  \includegraphics*[width=\columnwidth]{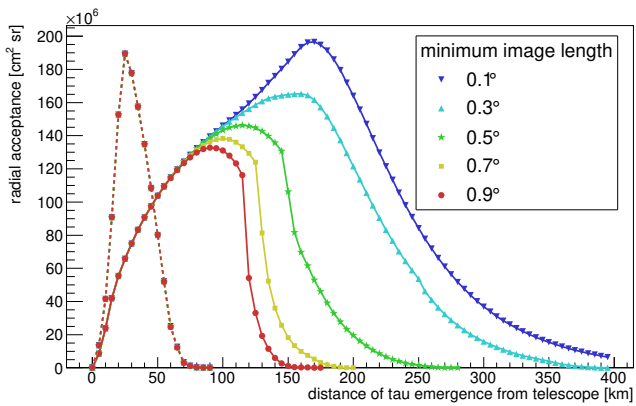}
\caption{
Radial acceptance for different required minimum image lengths
in the analysis. The dashed curves on the left give the acceptance of fluorescence detected
events. The solid curves on the right give the acceptance of Cherenkov detected
events.
}
  \label{fig:imagelength}
\end{figure}

We start by  discussing the impact of the minimum required image length in the
event reconstruction. Figure \ref{fig:imagelength} shows the acceptance as a
function of distance between the telescope and the point where the tau emerges
from the plane defined by $l$ and the horizon (see Fig.\ \ref{fig:geomview}).
The acceptance is calculated in 5\,km wide annuli centered at the position of
the telescope. Summing the acceptance of all bins yields the total acceptance
for tau neutrinos with energies from $10^{8.5}$\,GeV to $10^{9.5}$\,GeV. The
dashed curves on the left give the acceptance due to fluorescence detected
events while the solid curves on the right give the acceptance due to Cherenkov
detected events. 

As can be expected, the fluorescence emission is only detected up to a few tens
of kilometers from the telescope. Furthermore, the acceptance does not depend on
the minimum required image length because the showers develop so close that the
length is always above the required minimum. The situation is different for
Cherenkov detected events where the acceptance drops sharply above 100\,km if
the required image length is $>0.9^\circ$ but increases with a smaller required
image length.

Another observation is that fluorescence detected events dominate the acceptance
at small distances. This is because fluorescence detected events can be imaged
from all directions, i.e.\ also from the back, whereas Cherenkov detected
events can only be detected in a small range of $\alpha$. Overall, the
acceptance of fluorescence detected events is about 15\% of the acceptance of
Cherenkov detected events if a minimum image length of $0.1^\circ$ is required.
The fraction increases to  about 50\% for a minimum required image length of
$0.9^\circ$.

\begin{figure}[!tb]
  \centering
  \includegraphics*[width=\columnwidth]{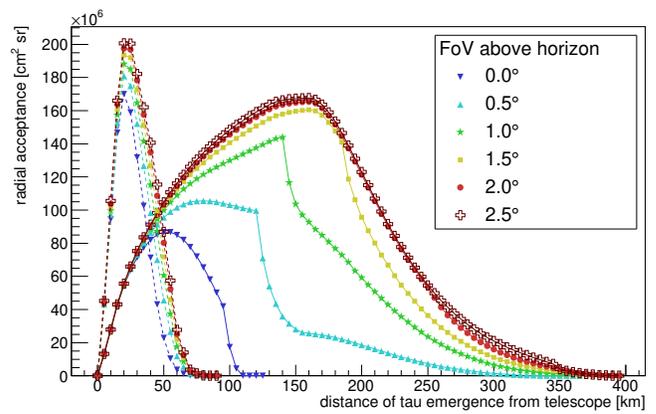}
\caption{
Radial acceptance for different vertical fields of view of the telescope camera
above the horizon. The camera field of view below the horizon is not limited.
The dashed curves on the left give the acceptance of fluorescence detected
events. The solid curves on the right give the acceptance of Cherenkov detected
events.
}
  \label{fig:UpperFoV}
\end{figure}

Figure \ref{fig:UpperFoV} shows the acceptance for six camera fields of views
above the horizon. The field of view below the horizon is not limited in these
calculations. While the acceptance due to fluorescence detected events shows
little dependence on the field of view above the horizon, it shows a strong
dependence for Cherenkov detected events. That is because taus that emerge far
away from the detector decay and produce an air shower in the camera field of
view that is above the horizon. The acceptance, however, does not increase significantly
if the field of view above the horizon is increased beyond $2^\circ$. This
feature can be used to define the field of view above $2^\circ$ as a veto and
thus reject potential background events due to cosmic rays.

\begin{figure}[!tb]
  \centering
  \includegraphics*[width=\columnwidth]{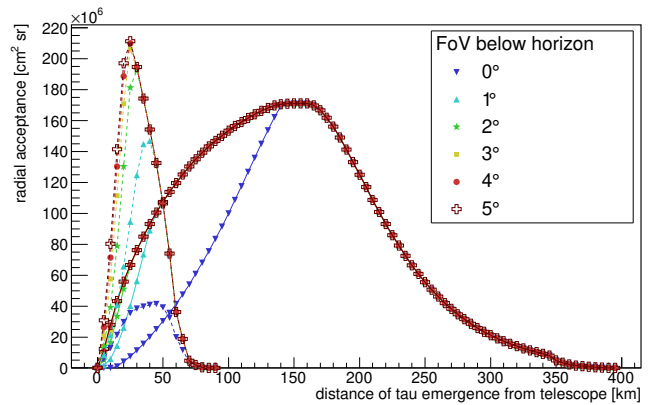}
\caption{
Radial acceptance for different vertical fields of view of the telescope camera
below the horizon. The camera field of view above the horizon is fixed at
$10^\circ$.
The dashed curves on the left give the acceptance of fluorescence detected
events. The solid curves on the right give the acceptance of Cherenkov detected
events.
}
  \label{fig:LowerFoV}
\end{figure}

The dependence of the acceptance on the camera field of view below the horizon
shows Fig.\ \ref{fig:LowerFoV}. Here we fix the field of view above the horizon
to $10^\circ$. As can be expected, the acceptance increases with an increasing
field of view for events that take place within 150\,km around the telescope,
which is the distance to the horizon. Increasing the field of view below the
horizon beyond $1^\circ$ does not impact the acceptance of Cherenkov detected
events as much as it does fluorescence detected events.  The majority of
fluorescence events is detected if the field of view extends to $2^\circ$ below
the horizon. 

\begin{figure}[!tb]
  \centering
  \includegraphics*[width=\columnwidth]{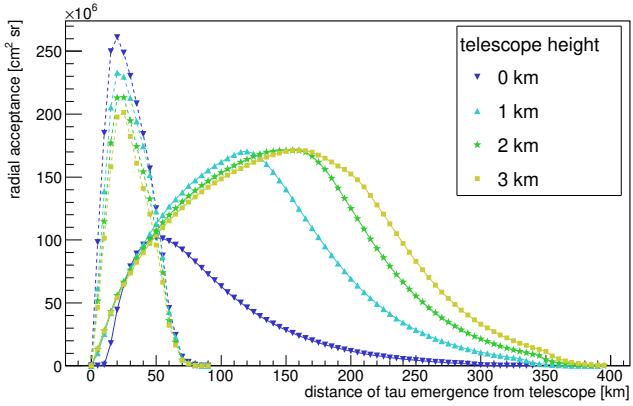}
\caption{
Radial acceptance for different locations of the telescope above ground. 
The dashed curves on the left give the acceptance of fluorescence detected
events. The solid curves on the right give the acceptance of Cherenkov detected
events.
}
  \label{fig:height}
\end{figure}

Figure \ref{fig:height} shows the acceptance for four telescope locations above
ground.  In order to not influence the calculation by a specific choice of the
camera field of view, we did not place a limit on the field of view below the
horizon and fixed the field of view above the horizon to $10^\circ$.  While the
acceptance of fluorescence detected events shows a slight decrease with
an increasing telescope height above ground, the acceptance of Cherenkov detected
events, which develop farther away from the telescope, clearly benefits from a
telescope location that is 1\,km or more above ground. That is because Cherenkov
light is less attenuated at higher altitudes.  For a location on the ground, the
acceptance of fluorescence detected events is about the same as it is for Cherenkov
detected events. When the telescope is placed 3\,km above the ground, about 15\%
of the total acceptance is due to fluorescence detected events.

\begin{figure}[!tb]
  \centering
  \includegraphics*[width=\columnwidth]{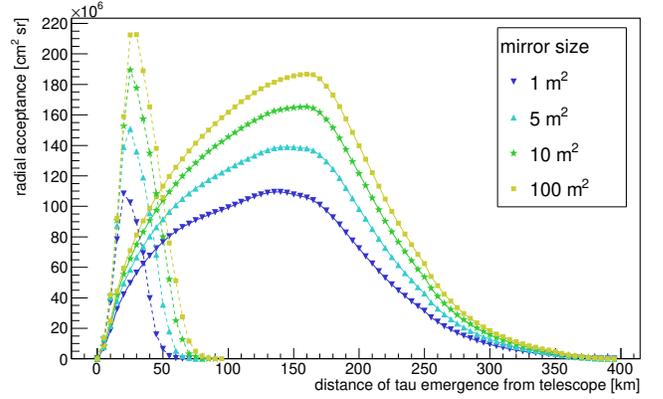}
\caption{
Radial acceptance for different effective mirror sizes. 
The dashed curves on the left give the acceptance of fluorescence detected
events. The solid curves on the right give the acceptance of Cherenkov detected
events.
}
  \label{fig:mirror}
\end{figure}

Figure \ref{fig:mirror} shows the acceptance for four different effective mirror
areas. A significant increase in acceptance is seen at all distances and for
both types of events as the mirror area increases from $1\,$m$^2$ to
$10\,$m$^2$.  Increasing the mirror area by a factor of 10 from 10\,m$^2$ to
100\,m$^2$, however, increases the acceptance by only an additional 13\%. 

\begin{figure}[!tb]
  \centering
  \includegraphics*[width=\columnwidth]{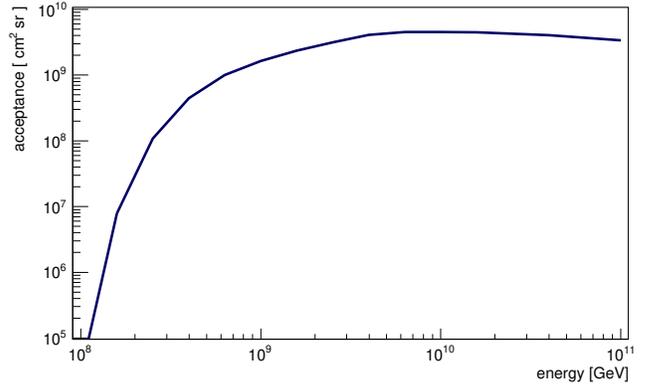}
\caption{
Acceptance of the baseline configuration. 
}
  \label{fig:baseaccept}
\end{figure}

\begin{figure}[!tb]
  \centering
  \includegraphics*[width=\columnwidth]{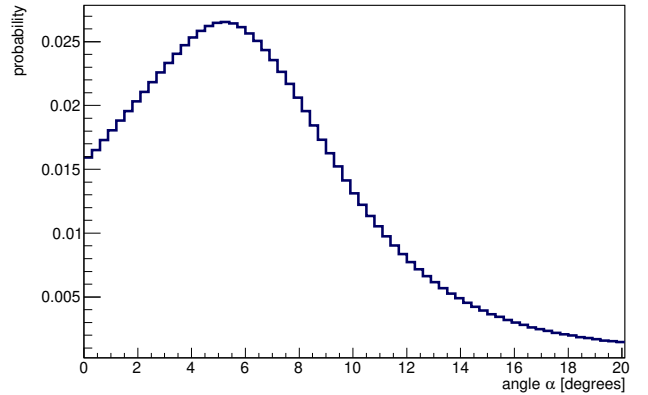}
\caption{
The $\alpha$ distribution of tau neutrinos with energies between
$10^7$\,GeV and $10^{11}$\,GeV that are detected with the baseline configuration.
}
  \label{fig:distranglealpha}
\end{figure}

We conclude our acceptance study with a calculation of the acceptance as
a function of energy for the baseline configuration from $10^7\,$GeV to
$10^{11}\,$GeV. The calculation shown in
Fig.\ \ref{fig:baseaccept} is done without integrating over neutrino energy;
i.e.\ it shows the acceptance as defined by Eq.\ (\ref{eq:1}). For that
same calculation, we show in Fig.\ \ref{fig:distranglealpha} the $\alpha$
distribution of events that trigger the readout. A total of 50\% of all detected events
have $\alpha<7^\circ$, and 90\% of all events have $\alpha<20^\circ$. The
remaining 10\% of the events detected at larger angles are fluorescence events.

\section{Sensitivity}

In this section we study how the sensitivity for $10^7$\,GeV to
$10^{10}$\,GeV tau neutrinos is affected by different telescope parameters. For
the baseline design we chose the same telescope configuration as in the previous
section, and we study the impact of one parameter at a time.

The sensitivity at a given energy is calculated by integrating the acceptance
over one decade of neutrino energies and centering the decade at that energy on a logarithmic
scale. For the flux of the neutrinos we again assume a power law with spectral index
$-2$.  We, furthermore, require one tau-neutrino detection in three years of
observation. The sensitivity is quoted for an all-flavor neutrino flux, which
assumes equal mixing into all three flavors by the time the neutrinos arrive at
Earth. 

For the duty cycle of the observation we assume 20\%, which is slightly more than the
typical duty cycle of a Cherenkov telescope (14\% or 1,200 hours of
observations). The slightly larger duty cycle is justified because observations
of tau neutrinos can also be carried out when clouds are present at high
altitudes. The actual duty cycle might in fact be larger because we assume that
only half of the time that is typically lost in Cherenkov telescope observations
due to bad weather can still be used for neutrino observations.  Tau-neutrino observations can also be
carried out at brighter moonlight than gamma-ray observations, which further
increases the duty cycle.

\begin{figure}[!tb]
\centering
\includegraphics*[width=\columnwidth]{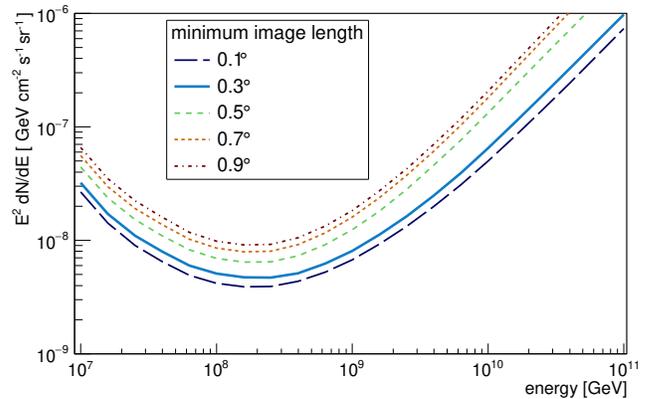} 
\caption{
Sensitivity integrated over one decade of energy for different minimal
shower image lengths required in the event reconstruction.
} 
\label{fig:sensitivityimage} 
\end{figure}

\begin{figure}[!tb]
  \centering
  \includegraphics*[width=\columnwidth]{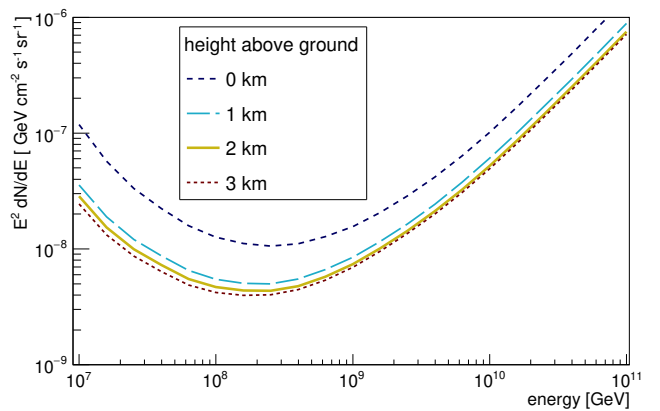}
\caption{
Sensitivity integrated over one decade of energy for four locations of
the telescope above ground.
  \label{fig:sensitivityheight}
}
\end{figure}

\begin{figure}[!tb]
  \centering
  \includegraphics*[width=\columnwidth]{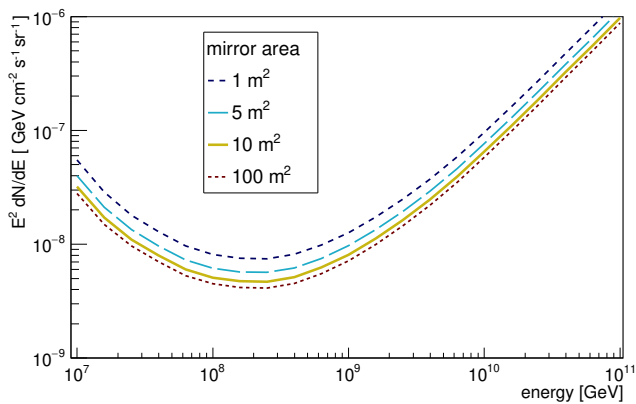}
\caption{
Sensitivity integrated over one decade of energy for four effective
mirror sizes.
}
  \label{fig:sensitivitymirror}
\end{figure}

\begin{figure}[!tb]
  \centering
  \includegraphics*[width=\columnwidth]{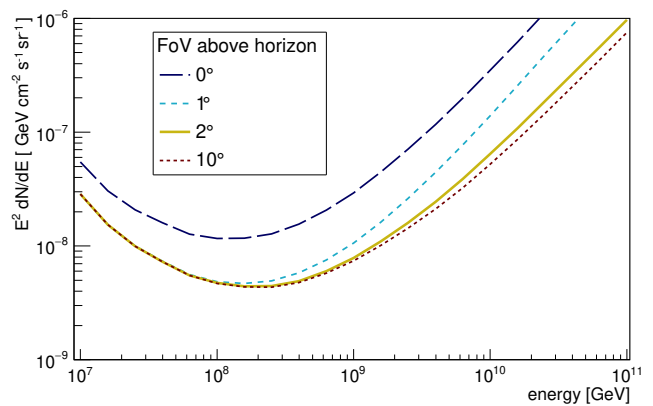}
\caption{
Sensitivity integrated over one decade of energy for four fields of
views above the horizon. The field of view below the horizon is maximal.
}
  \label{fig:sensitivityUpperFoV}
\end{figure}

\begin{figure}[!tb]
  \centering
  \includegraphics*[width=\columnwidth]{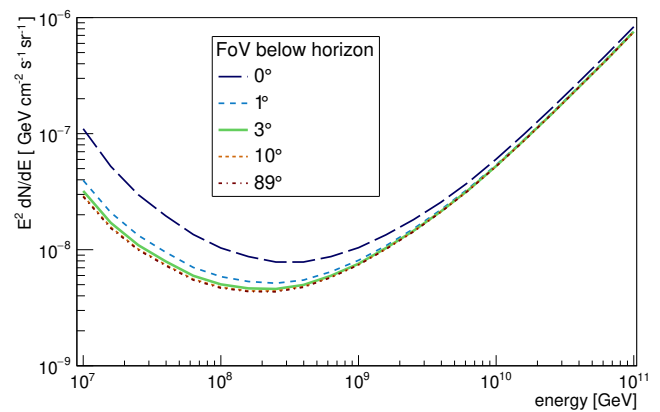}
\caption{
Sensitivity integrated over one decade of energy for five fields of
views below the horizon. The field of view above the horizon is fixed at
$10^\circ$.
}
  \label{fig:sensitivityLowerFoV}
\end{figure}

\begin{figure}[!tb]
  \centering
  \includegraphics*[width=\columnwidth]{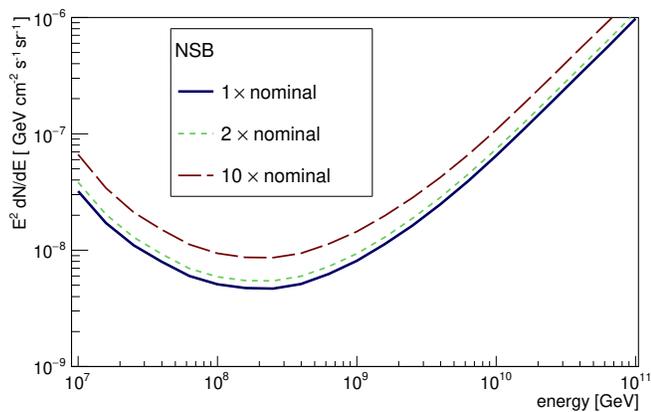}
\caption{
Sensitivity integrated over one decade of energy for three levels of night-sky
background in multiples of a nominal night-sky background of
$3.7\times10^6$\,photoelectrons/s/mm$^2$/sr.
}
  \label{fig:sensitivityNSB}
\end{figure}

The results of the different sensitivity calculations are shown in Figs.\
\ref{fig:sensitivityimage}--\ref{fig:sensitivityNSB}. The solid line in each
figure depicts the sensitivity of the baseline configuration.

All sensitivity curves have in common that the best sensitivity is assumed
between $2\times10^8$\,GeV and $4\times10^8$\,GeV. The worsening of the
sensitivity at high energies is explained with the increasing tau decay length.
As a result of the increasing decay length, a decreasing fraction of high-energy
taus decay and develop a full shower within the field of view of the telescope.
In our calculations we neglect that some events could be viewed from the back if
the tau passes across the telescope and decays within the field of view of the
telescope opposite from where it originated.

 The worsening of the sensitivity at lower energies is
mostly due to a decrease in the emergence probability and a reduced light
intensity of the air shower if the tau emerges and produces a shower.  

Figure \ref{fig:sensitivityimage} shows the sensitivity for different required
minimum image lengths in the reconstruction. The sensitivity improves if smaller
images can be reconstructed because showers developing farther away from the
telescope are accepted in the reconstruction. The sensitivity in the core energy
region improves by a factor of 2 if the smallest reconstructable image length
changes from $0.9^{\circ}$ to $0.1^{\circ}$.

Figure \ref{fig:sensitivityheight} shows the sensitivity for different telescope
locations above the ground. In order to not be affected by the choice of field
of view, the upper edge of the field of view is set to $10^\circ$ above the
horizon and the lower edge to $90^\circ$ below the horizon.  The most dramatic
sensitivity improvements occur when the telescope is moved from the ground to a
height of 1\,km, which reflects the increase in acceptance to distant showers
due to reduced atmospheric absorption (Fig.\ \ref{fig:height}). Increasing the
height above ground by an additional 2\,km to 3\,km improves the sensitivity by
another 20\%. The conclusion to draw is that the instrument is best located at an
altitude between 1\,km and 3\,km above ground.

Figure \ref{fig:sensitivitymirror} shows the sensitivity for different effective
mirror sizes. Unlike in the previous two cases where about the same
relative change in sensitivity is observed at all energies, an increased mirror
has a bigger impact at lower energies than at higher energies. That is
because dimmer showers become detectable at lower energies while the majority of
the showers at higher energies are already detectable with a smaller
mirror.  Increasing the light collection area thus improves the sensitivity at
all energies but shifts the energy where the best sensitivity is obtained toward smaller energies.

An interesting observation is that the sensitivity improves by a factor of $1.6$
when the mirror area increases from 1\,m${^2}$ to 10\,m${^2}$ but only by a
factor of $1.2$ when the mirror increases another factor of 10 from 10\,m${^2}$
to 100\,m${^2}$. We interpret this as evidence that neutrinos, which produce an
emerging tau and a subsequent particle shower are already detected with almost
100\% efficiency with a 10\,m${^2}$ mirror surface. What we do not
take into account here is that a large mirror captures fainter events and more
details of shower images, which can potentially be reconstructed with a finer
pixelated camera and thus could increase the sensitivity of a 100\,m${^2}$ class
instrument beyond what our study indicates. 

How the field of view of the camera above and below the horizon affects the
sensitivity is shown in Figs.\ \ref{fig:sensitivityUpperFoV} and
\ref{fig:sensitivityLowerFoV}, respectively. In Fig.\
\ref{fig:sensitivityUpperFoV} the field of view below the horizon is fixed to
$89^\circ$, the maximum possible value, and the field of view above the horizon
is varied between $0^\circ$ and $10^\circ$. At energies below $10^8$\,GeV, the
sensitivity reaches its best value already for a field of view above the horizon
of $1^\circ$. At higher energies the sensitivity improves
significantly up to a $2^\circ$ field of view above the horizon and continues to
improve by only about 20\% with a further increase to
$10^\circ$.

In Fig.\ \ref{fig:sensitivityLowerFoV}, the field of view above the horizon is
fixed at $10^\circ$ and varied between $0^\circ$ and $89^\circ$ below the
horizon. Increasing the
field of view below the horizon to $3^\circ$ improves the sensitivity almost to the best possible
sensitivity, which is achieved if the field of view below the horizon is
$89^\circ$. The effect of changing the field of view below the horizon is less
pronounced at higher energies, which is to be expected because the majority of
detectable high-energy showers originate at large distances from the telescope
and develop a shower above the horizon. In other words the location of a
recorded image in the camera contains information about the energy of the tau
neutrino.

Combining the results from varying the field of view above and below the horizon
we conclude that a camera with a vertical field of view of $5^\circ$ and a
telescope pointing at the horizon allows one to image most events if the telescope
is located 2\,km above ground. This leaves $0.5^\circ$ at the top of the field
of view, which can be used as a veto.  Images that are partially located
in the veto region are likely due to downward going showers, meteorites, or
lightning and can thus be rejected without affecting the acceptance of
neutrinos.

The last effect we studied is the impact of different night-sky background
levels on the sensitivity. Figure \ref{fig:sensitivityNSB} shows the sensitivity
of the baseline configuration for the nominal NSB intensity of
$3.7\times10^6$\,photoelectrons/s/mm$^2$/sr as well as for 2 and for 10 times the nominal value. The pixel trigger thresholds in increasing
order of the NSB levels are 24\,pe, 41\,pe, and 155\,pe and in each case result
in a telescope trigger rate of 1\,Hz due to NSB fluctuations. The
sensitivity worsens with increasing NSB by about the same factor
at all energies. For 2 times the NSB, the sensitivity worsens by
15\%, and for 10 times the nominal NSB the sensitivity worsens by 85\%. Because of the higher trigger
threshold, events farther away from the telescope and those viewed at
larger angles $\alpha$ do not trigger the readout anymore. But because the
Cherenkov intensity changes exponentially with distance and $\alpha$, the effect
of a higher trigger threshold on the sensitivity is only logarithmic. 

\section{Trinity Design Considerations}

We now discuss how the above results translate into design requirements for an
actual instrument, which we call \emph{Trinity}.  The basis of our study is an
imaging detector with a $360^\circ$ azimuthal field of view and a location on
top of a mountain. 

The required angular resolution of the optics across the field of view is quite
modest and can be a few tenths of a degree as long as it resolves shower images
above the minimum image length.  Several optics concepts exist, which satisfy
that requirement. A Schmidt optics, as it is used in the fluorescence detectors
of TA and AUGER, is an obvious good choice
\cite{Abraham2010,Abbasi2015,Tokuno2012}.  Another option, which yields a larger
azimuthal field of view with a single optical system and thus only requires six
stations, is the $60^\circ\times5^\circ$ MACHETE optics \cite{Cortina2016}.
Based on Fig.\ \ref{fig:sensitivitymirror}, the sensitivity approaches its
optimum for an effective mirror area of 10\,m$^2$, which translates into a
$4.1$\,m$\times7.9$\,m  mirror area of one MACHETE station.  

The camera in the focal plane of the optics records the shower image. It
consists of the photon detectors and the readout electronics. As already
discussed, SiPMs have a spectral response, which is a good match to the
Cherenkov spectrum after atmospheric absorption at the telescope. For an angular
resolution of $0.3^{\circ}$ one MACHETE camera \cite{Cortina2016} would have
3,333 pixels, each with a size of $19$\,mm$\times19$\,mm. A pixel would
consist of a nonimaging light concentrator coupled to a $9$\,mm$\times9$\,mm
SiPM. 

The spread in the photon arrival times is a main driver defining the minimum
sampling speed for the digitization of the SiPM signals. In Cherenkov
telescopes,
the shower is typically imaged under a small angle $\alpha<2^\circ$. For such
small angles, the arrival-time distribution is on the order of nanoseconds and
sampling speeds up into the gigasamples per second territory help to keep the
contamination of the Cherenkov signal from background photons at a minimum.  The
fast sampling also helps in the reconstruction of events \cite{Aliu2009a}. 
Fluorescence detectors, on the other hand, require sampling speeds of only tens
of megasamples per second to capture a signal, which is spread out over several
microseconds \cite{Gemmeke2000a}. For an imaging system dedicated to the
detection of tau neutrinos, a sampling speed of about 100 megasamples
per second should be sufficient because the most probable value for $\alpha$ is
$\sim5^\circ$ (see Fig.\ \ref{fig:distranglealpha}) and the Cherenkov photons
for that viewing angle arrive within $10$\,ns\ $\cdots100$\,ns (see Fig.\
\ref{fig:ArrivalTimeSpreadvsAzimuth}).

A preferred solution for the readout would be the concept used in AUGER or
TA. There, the signals are continuously digitized, and the trigger decision and
signal processing are accomplished digitally \cite{Abraham2010,Tameda2009,Gemmeke2000a}. An alternative
readout concept would use a switch-capacitor array that samples the signals with 100
megasamples per second. The AGET system developed for time-projection chambers
could be a viable option \cite{Pollacco2018}.

\section{Discussion}

The observation of ultrahigh-energy neutrinos has the potential to offer some
unique insight into long-standing questions in astrophysics and neutrino
physics. It is, therefore, not surprising that a number of experiments are
proposed to explore the UHE neutrino band. In this paper we took a closer look
at the possibility of detecting earth-skimming tau neutrinos with
\emph{Trinity},
an imaging system located on top of a mountain. 

We find that the energy dependence of \emph{Trinity}'s sensitivity is mostly
driven by the interaction cross section of tau neutrinos and the tau decay
length, with the best sensitivity being reached at $2\times10^8$\,GeV. Among the
different telescope configurations tested, one that achieves a good compromise
between performance and cost has an effective mirror area of 10\,m$^2$ and is
located 2\,km above the ground. If the angular resolution of the camera and the
optics of such a system is good enough to allow the reconstruction of shower
images as small as $0.3^\circ$, a sensitivity of
$5\times10^{-9}$\,GeV\,cm$^{-2}$\,s$^{-1}$\,sr$^{-1}$ per decade energy is
achievable.

\begin{figure}[!tb]
  \centering
  \includegraphics*[width=\columnwidth]{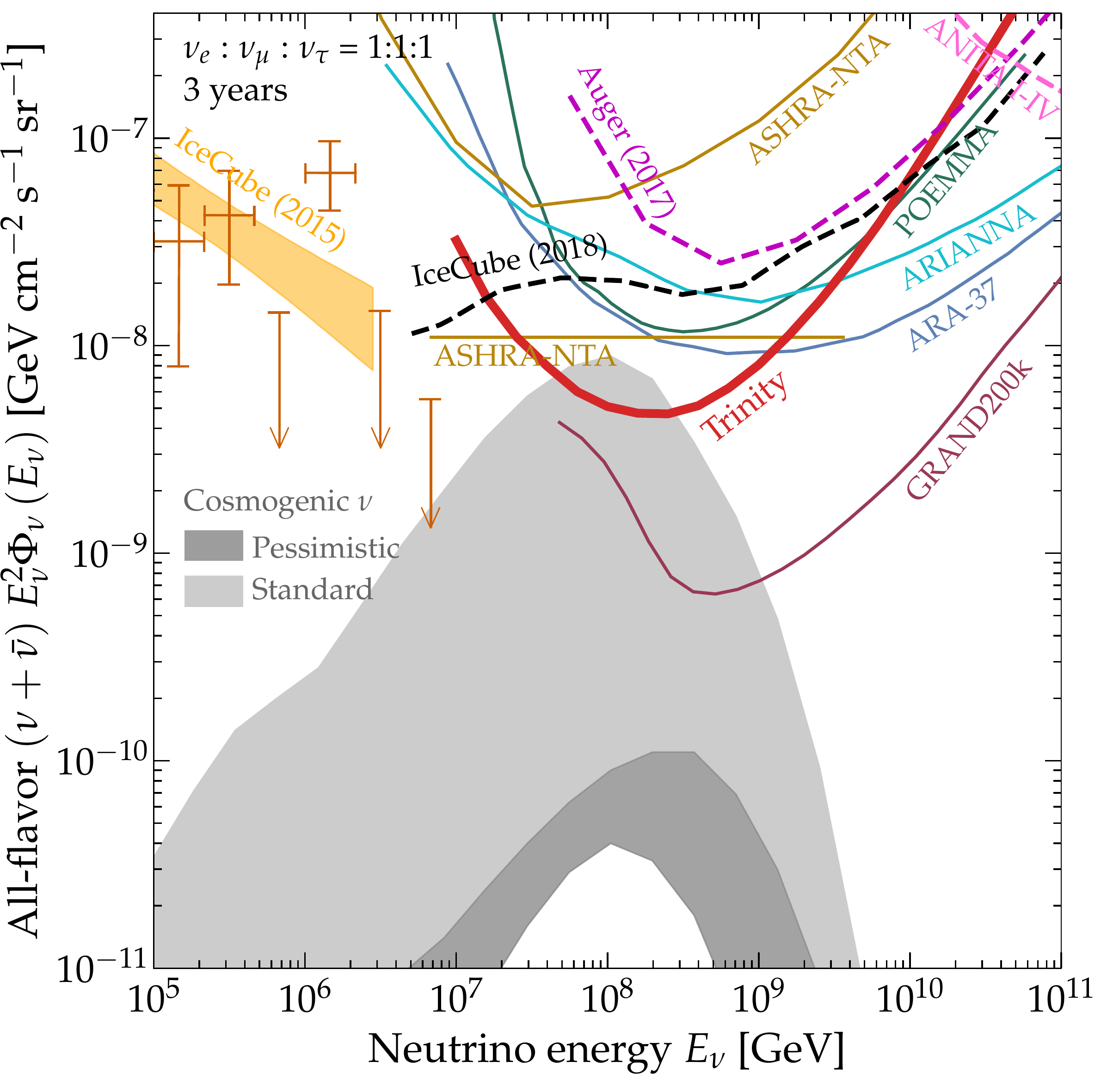}
\caption{
Sensitivity of \emph{Trinity} and other experiments, neutrino flux
predictions, and existing flux constraints. Figure adapted from
\cite{GRANDCollaboration2018}.
}
  \label{fig:sensitivityAllExperiments}
\end{figure}

A comparison of \emph{Trinity}'s sensitivity with that of GRAND \cite{GRANDCollaboration2018},
POEMMA \cite{Krizmanic2018}, ARIANNA \cite{Persichilli2018}, ARA-37 \cite{Allison2016}, and NTA
\cite{Sasaki2014}
is shown in Fig.\ \ref{fig:sensitivityAllExperiments}. Also shown are predictions of
the cosmogenic neutrino flux \cite{Batista2018}, measurements of the
astrophysical neutrino flux with IceCube \cite{IceCubeCollaboration2015}, and the most
recent limits from IceCube \cite{Aartsen2018}, AUGER
\cite{ThePierreAugerCollaboration2017}, and ANITA \cite{Gorham2019} . At energies where
it has the most sensitivity, \emph{Trinity} delivers a sensitivity that falls in between the sensitivity of proposed in-ice radio detectors
and GRAND. The sensitivity is sufficient to probe an extension of the IceCube
detected astrophysical neutrino flux and predictions of the cosmogenic neutrino
flux. 

\emph{Trinity} is complementary to in-ice radio detectors, which are sensitive
to all neutrino flavors. Comparing fluxes measured with both techniques allows
one to do neutrino physics at the highest energies. The very different detection
techniques would also allow one to do systematic cross-checks. These
cross-checks could even take place at the event level in case \emph{Trinity} and
an earth-skimming radio detector like GRAND would be deployed at the same site.

Given the expected  moderate costs of \emph{Trinity}, it is feasible to scale
the system by deploying several \emph{Trinity} sized detector stations in both
hemispheres at different locations. The combined sensitivity to a diffuse
neutrino flux of such a distributed system scales proportional to the number
of detector stations. Furthermore, complementary parts of the sky would be
observed, thus providing a better coverage in searches for astrophysical
neutrino sources. It is also possible to build \emph{Trinity} not on one site
with a $360^\circ$ azimuthal field of view but to deploy the six detector stations
on different sites, each monitoring a different field of view and thus gaining
the same sensitivity.

Because the neutrino interaction cross section and the tau decay length are
dominant factors determining the sensitivity, we are open to the idea that other
topographies than the one studied here could yield a higher sensitivity.  In
particular, a mountain range with the right height and width, as well as the
right distance to the telescope would increase the solid angle acceptance as is
evidenced in \cite{Gora2015}.

For NTA it is proposed to observe taus appearing from a volcano on Mauna Loa
\cite{Sasaki2014}. Despite the very different topography, the integral
sensitivity predicted for NTA is comparable to \emph{Trinity}'s sensitivity (see
Fig.\
\ref{fig:sensitivityAllExperiments}), which is further evidence that other
topographies are preferable to the one we studied here. An NTA-like topography has
additional operational advantages. Air showers, for example, develop much closer to
the telescope and the angular size of an image is, therefore, larger, which
allows for a coarser angular resolution of the camera, i.e.\ fewer
readout channels and thus a reduction in costs. The closer proximity to
the shower  also increases the light intensity, thus allowing for smaller
mirrors.  Another advantage is reduced atmospheric absorption and an easier
monitoring of the air mass, which reduces systematic uncertainties in the energy
reconstruction.

The biggest uncertainty in our sensitivity calculations comes from the
simplified trigger simulation and the assumption that the measurement is
background free.  For the trigger, we did not investigate how much the spread in
arrival times for larger angles $\alpha$ affects the trigger efficiency or
what the impact of different trigger strategies is on the sensitivity. A perhaps
more efficient, but also more complicated, trigger strategy than the one we
studied would take the time gradient across the shower image into account. 

The assumption of being background free implies that only images of tau
initiated air showers survive the event reconstruction and other events
triggering the readout are rejected in the analysis.  Potential background
events are due to fluctuations in the NSB, cosmic-ray events, and isolated
muons. Background events due to fluctuations in the NSB are reliably suppressed
by the standard principal component analysis of air-shower images. Cosmic-ray
air showers at large zenith angles develop at much larger distances. Their light
is, therefore, subject to more absorption and scattering. Cosmic-ray showers,
furthermore, start above the horizon, whereas tau showers start below the
horizon, which can be distinguished based on the time and spatial
characteristics of the recorded shower images, and their location in the camera.
Isolated muons could be another source of background events, which becomes
important for larger mirror surfaces than considered here. Muon events are
easily suppressed by operating two detector stations separated by more than
100\,m.  They can also be rejected based on the much narrower photon
arrival-time distribution of only a few nanoseconds compared to several tens to
hundreds of nanoseconds for a tau initiated shower.  How well these background
events can be suppressed remains to be seen and is best assessed with
observations.

We conclude that the imaging of air showers is a viable technique to detect tau
neutrinos. Combined with recent advances in the SiPM technology and affordable
readout options with sampling speeds in the 100\,MS/s range, a
\emph{Trinity}-like detector is a cost effective and robust way to detect tau
neutrinos with competitive sensitivity.

\section*{Acknowledgments} 
This project was started by a discussion with Albrecht Karle at VHEPA in 2016.
I am grateful for the many discussions I had with him and a number of other
colleagues over the past three years.  Mary Hall Reno provided insight into the
calculations of the tau emergence probabilities. Juan Cortina provided insight about the
expected optical resolution of a scaled MACHETE. Abelardo Moralejo and Michael
Unger gave valuable feedback about the shower and Cherenkov light
parametrization and the detector configuration.  Mauricio Bustamante was so
kind to provide me with Fig.\ \ref{fig:sensitivityAllExperiments}. Dieter
Heck's relentless support has proved vital in my effort to set up the CORSIKA
simulations and to understand the simulation results.

\section*{References}

\bibliographystyle{apsrev4-1}
\bibliography{GZKNeutrino}

\end{document}